\documentclass[
 reprint,
superscriptaddress,
nofootinbib,
 amsmath,amssymb,
 aps,pra,
]{revtex4-1}
\usepackage{amsmath}
\usepackage{graphicx}
\usepackage{dcolumn}
\usepackage[normalem]{ulem}
\usepackage[colorlinks=true,linkcolor=blue,urlcolor=blue,citecolor=blue]{hyperref}

\begin{document}

\title{Symmetry breaking exhibition by magnetic field induced explicit circular dichroism}

\author{A. Tonoyan}
\affiliation{Institute for Physical Research, NAS of Armenia, Ashtarak-2, 0203 Armenia}
\affiliation{Laboratoire Interdisciplinaire Carnot de Bourgogne, UMR CNRS 6303 - Universit\'e Bourgogne - Franche-Comt\'e, BP 47870, 21078 Dijon Cedex, France}
\author{A. Sargsyan}
\affiliation{Institute for Physical Research, NAS of Armenia, Ashtarak-2, 0203 Armenia}
\author{E. Klinger}
\affiliation{Institute for Physical Research, NAS of Armenia, Ashtarak-2, 0203 Armenia}
\affiliation{Laboratoire Interdisciplinaire Carnot de Bourgogne, UMR CNRS 6303 - Universit\'e Bourgogne - Franche-Comt\'e, BP 47870, 21078 Dijon Cedex, France}
\author{G. Hakhumyan}
\affiliation{Institute for Physical Research, NAS of Armenia, Ashtarak-2, 0203 Armenia}
\author{C. Leroy}
\affiliation{Laboratoire Interdisciplinaire Carnot de Bourgogne, UMR CNRS 6303 - Universit\'e Bourgogne - Franche-Comt\'e, BP 47870, 21078 Dijon Cedex, France}
\author{M. Auzinsh}
\affiliation{Department of Physics, University of Latvia, 19 Rainis Blvd., Riga LV-1586, Latvia}
\author{A. Papoyan}
\affiliation{Institute for Physical Research, NAS of Armenia, Ashtarak-2, 0203 Armenia}
\author{D. Sarkisyan}
\affiliation{Institute for Physical Research, NAS of Armenia, Ashtarak-2, 0203 Armenia}

\date{\today}

\begin{abstract}
In this letter we demonstrate universal symmetry breaking by means of magnetically induced circular dichroism. Magnetic field induces forbidden at zero field atomic transitions between $\Delta F = \pm2$ hyperfine levels. In a particular range of magnetic field, intensities of these transitions experience significant enhancement. We have deduced a general rule applicable for the $D_2$ lines of all bosonic alkali atoms, that is transition intensity enhancement is larger for the case of $\sigma^+$ than for $\sigma^-$ excitation for $\Delta F = +2$, whereas it is larger ($\textit{e.g.}$ up to $10^{11}$ times for $^{85}$Rb atoms) in the case of $\sigma^-$ than for $\sigma^+$ polarization for $\Delta F = -2$. This asymmetric behaviour results in an explicit circular dichroism. For experimental verification we employed half-wavelength-thick atomic vapor nanocells using a derivative of selective reflection technique, which provides sub-Doppler spectroscopic linewidth ($\sim$50 MHz). The presented theoretical curves well  describe the experimental results. This effect can find applications particularly in parity violation experiments.

\end{abstract}

\maketitle

 In the presence of external magnetic field magnetic circular dichroism (MCD) in an atomic media for right ($\sigma^-$) and left ($\sigma^+$) circularly polarized radiation attracts growing attention 
due to numerous applications. This process is intrinsically linked with magnetic birefringence, which underlies linear and nonlinear Faraday rotation. Magneto-optical processes are successfully used in magneto-optical tomography, narrowband atomic filtering, optical magnetometry, tunable laser frequency locking, etc. \cite{bud1,yas1,auz1,bi1,zen1,budker,katso}. Fundamental interest towards MCD is particularly connected with studies of parity violation in optical experiments with heavy atoms \cite{ale1,bou1,wag1,koz1,giena}. 

A strong MCD has been measured for the number of electrons ejected by ionization of atomic helium using right-polarized light \cite{ilc1}. Magnetic field-induced modification of atomic transitions probabilities of $F_g=3 \rightarrow F_e=5$ of Cs $D_2$ line, which are forbidden at zero magnetic field was reported for $\sigma^+$ polarized laser radiation \cite{sar1}.

\begin{figure}[b]
	\begin{center}
		\includegraphics[width=200pt]{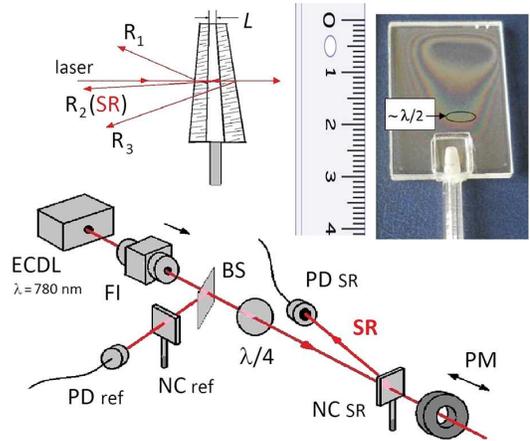}
		\caption{\label{fig:fig1} Layout of the experimental setup: (ECDL) cw laser with $\lambda$ = 780 nm, (FI) Faraday isolator, (NC) Rb nanocell inside an oven, (PM) permanent magnet, (PD) photodetector, ($\lambda$/4) quarter-wave plate ("SR" and "ref" stand for selective reflection and reference channels, respectively). Upper right inset: photograph of the NC, oval marks region around 390 nm. Upper left inset: geometry of the three reflected beams. The beam (SR) propagates in the direction of R$_2$.}
	\end{center}
\end{figure}

Previously we have shown that in a particular range of external magnetic field that depends on hyperfine splitting of an isotope, $D_2$ lines of all alkali atoms contain two groups of transitions which are forbidden by the $\Delta F=0, \pm 1$ selection rule for the total angular momentum at zero magnetic field \cite{sar1}. Application of a magnetic field causes mixing of states, allowing transitions between the states with $\Delta F=\pm 2$, which we call  magnetically induced (MI). For the case of Rb these transitions are: $F_g=2 \rightarrow F_e=4$ and $F_g=3 \rightarrow F_e=1$ for the $^{85}$Rb isotope, and $F_g=1 \rightarrow F_e=3$ and $F_g=2 \rightarrow F_e=0$ for the $^{87}$Rb isotope. The transitions obeying the selection rules $\Delta F=+2$ and $\Delta F=-2$ in the first and second groups correspondingly,  behave distinctly depending on whether the excitation radiation is $\sigma^+$ or $\sigma^-$ polarized, more importantly, this relates also to the individual transitions in each group. We show that for the first group ($\Delta F=+2$) of $^{85}$Rb, the strongest transition probability for the case of $\sigma^+$ is almost five times higher than that for $\sigma^-$ polarization, while for the second group ($\Delta F=-2$) the strongest transition probability in the case of $\sigma^-$ is about $10^{11}$ times higher than that for $\sigma^+$ polarization. In other words, in the first group for the case of $\sigma^+$ radiation, transitions are stronger than in the case of $\sigma^-$, whereas in the second group the transitions are stronger for the case of $\sigma^-$ polarization. This difference leads to an explicit circular dichroism.

\begin{figure*}[t]
\begin{center}
\includegraphics[width=0.8\textwidth]{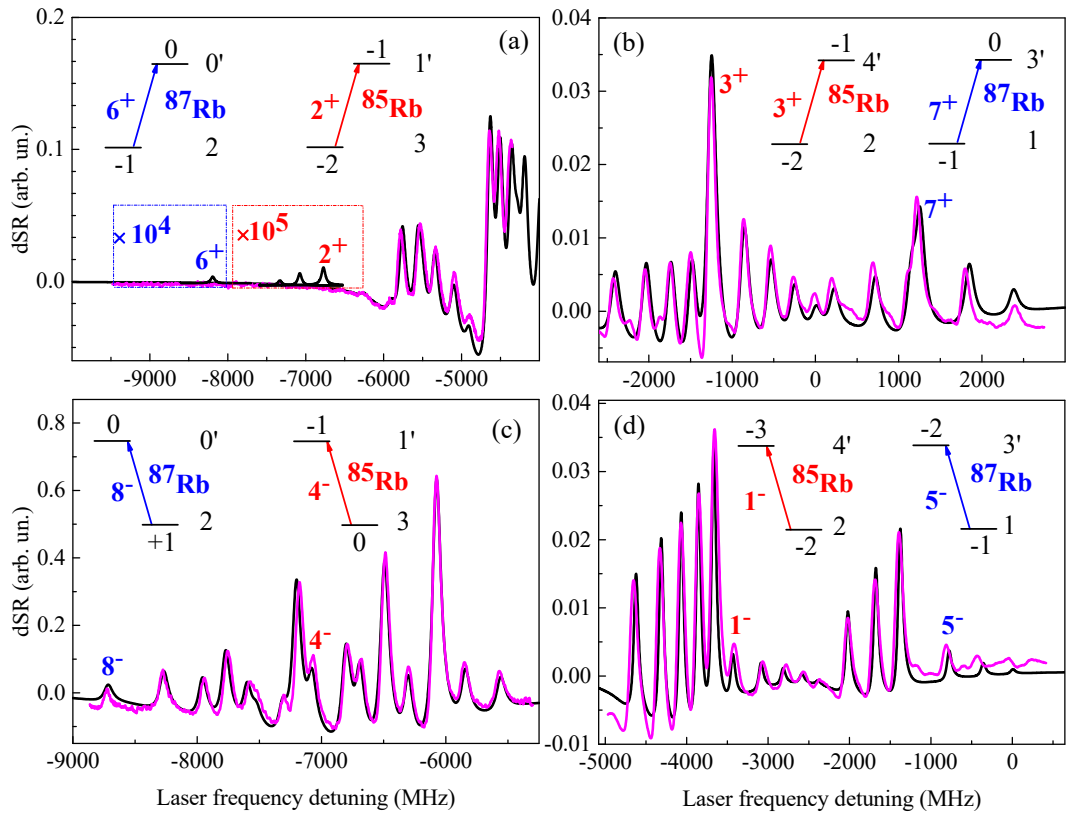}
  \caption{\label{fig:fig2} The experimental (magenta line) and theoretical dSR (black line) spectra of the Rb vapour in nanocell with the length of 390 nm. Peaks are labelled according to the diagrams. The low frequency part of the spectra for (a) $\sigma^+$ and (c) $\sigma^-$ polarization at 500 G; the high frequency part of the spectra for (b) $\sigma^+$ and (d) $\sigma^-$ at 650 G.}
\end{center}
\end{figure*}

In the present work we studied, both experimentally and theoretically, an explicit circular dichroism for MI transitions of $^{85}$Rb and $^{87}$Rb $D_2$ lines. The observed behavior is further generalized for the $D_2$ line transitions of all alkali atoms. The studies have been performed by derivative of selective reflection (dSR) technique using a Rb vapor nanocell with half resonance wavelength thickness ($L \approx \lambda/2 =$ 390 nm), which provides strong narrowing of spectral lines (40 - 50 MHz)  and uniformity of the applied $B$-field in the interaction region \cite{sar2,sar3}.

The experimental arrangement is schematically sketched in Fig.~\ref{fig:fig1}.   A single-frequency $\gamma_L/2\pi \approx$ 1~MHz bandwidth cw external cavity diode laser with $\lambda=780$~nm wavelength is used. The collimated $P_L \approx 20~\mu$W circularly polarized beam is directed at normal incidence onto the   $L = \lambda/2$ area of the Rb nanocell's (NC). The general design of NC is similar to that described in \cite{kea1,whi1,kea2}. A compact oven is used to maintain the required temperature regime ($T_R \approx 100~^\circ$C for thin sapphire reservoir containing metallic Rb, and about 20~$^\circ$C higher at the windows) to obtain $\sim$ 10$^{13}$ cm$^{-3}$ atomic number density.
Smooth vertical translation of the cell and oven assembly allows an appropriate adjustment of the atomic vapor column thickness avoiding the variation of thermal conditions.

\begin{figure}[ht]
\begin{center}
\includegraphics[width=0.5\textwidth]{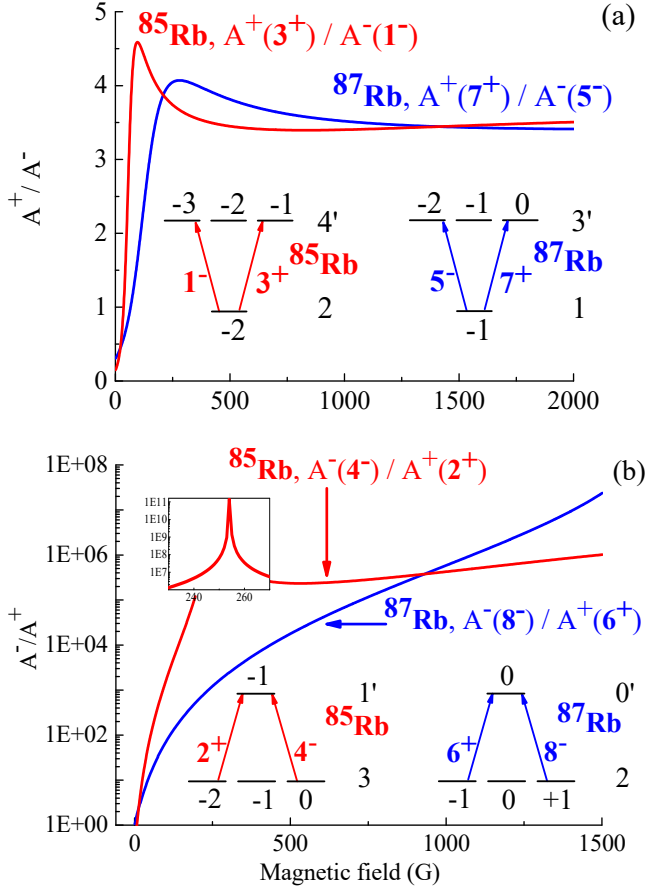}
\caption{\label{fig:fig3} Theoretical $B$-field dependencies of the ratio of MI transition probabilities. Red and blue lines are for $^{85}$Rb and $^{87}$Rb respectively. a) $A^+(3^+)/A^-(1^-)$ and $A^+(7^+)/A^-(5^-)$ (first group). b) $A^-(4^-)/A^+(2^+)$ and $A^-(8^-)/A^+(6^+)$ (second group).The diagrams of corresponding MI transitions of $^{85}$Rb and $^{87}$Rb are shown in the insets.}
\end{center}
\end{figure}

The selective reflection (SR) beam is separated from two other beams reflected from the front and rear surfaces of the cell (see the upper left inset of Fig. \ref{fig:fig1}), and after passing through narrow-band $\lambda=$ 780 nm filter is recorded by a photodiode (PD$_{SR}$). 
A strong permanent Nd magnet mounted on the micrometric-step translation stage in the proximity of the cell's rear window provides calibrated magnetic field.  The $B$-field strength is varied by simple longitudinal displacement of the magnet. To have a reference signal, a fraction of laser radiation is directed to an auxiliary $L = \lambda$ Rb nanocell NC$_{ref}$ \cite{sar4}. The reference signal is recorded simultaneously with the SR signal while the laser frequency is linearly scanned across the $D_2$ line. The scanning rate is chosen to be slow enough for assuring a steady-state interaction regime.

\begin{figure}[ht]
\begin{center}
\includegraphics[width=0.45\textwidth]{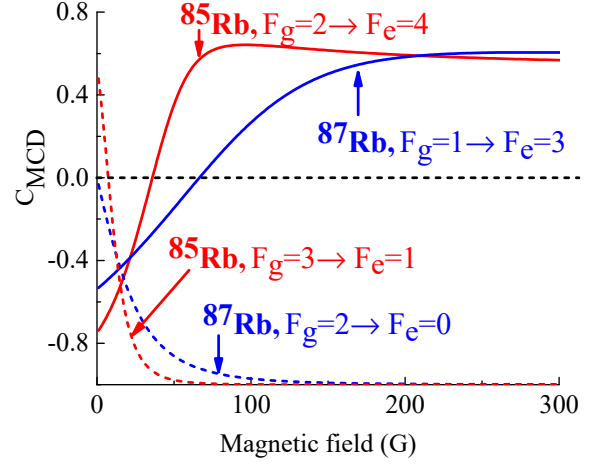}
\caption{\label{fig:fig4} Magnetic field dependence of MCD figure of merit coefficient $C_{MCD}$ for different magnetic-induced transitions of the Rb.}
\end{center}
\end{figure}

One of the advantages of our experimental setup is its simplicity, it allows not to use complex setup with vacuum chamber to produce cold atoms in order to have light-atom interaction for the same velocity atomic group. Vertically (to light beam) flying atoms in a nanocell provide narrow spectral lines. Secondly, because of the high temperature ($\sim$100~$^\circ$C), it is possible to have sufficient number of atoms in light-atom interaction area to have a relevant dSR spectrum. And lastly, in our setup it is needed only one and low power cw laser beam, which allows to avoid saturation effects. 

The obtained complete spectra of $D_2$ lines of Rb for  $\sigma^+$ and $\sigma^-$ polarizations are shown in Fig.~\ref{fig:fig2}a,b and Fig.~\ref{fig:fig2}c,d, respectively. 
While Figs. \ref{fig:fig2}a and \ref{fig:fig2}c represent the low frequency parts of the $D_2$ lines containing transitions of $\Delta F=-2$ group at 500 G, Fig.~\ref{fig:fig2}b and \ref{fig:fig2}d represent the high frequency part of the $D_2$ lines containing transitions of $\Delta F=+2$ group at 650 G. Labeled are only the strongest eight out of 24 MI transitions and the labeling on the plots corresponds to the transition diagrams in insets. Magenta line represents the experiment and the black line the theory, the latter is based on the model described in detail in Ref. \cite{tre1,dut1} and the mismatch of the theory  and experiment is less than 5$\%$. At zero and $B\gg B_0$ ($B_0 = A_{hfs}/ \mu _B$, $A_{hfs}$ is the hyperfine coupling constant of ground states and $\mu _B$ is the Bohr magneton \cite{ols1}) magnetic fields the intensities of the labelled transitions are zero, while in the presence of intermediate magnetic field, as we see, some of them are strongly enhanced. From the comparison of Fig.~\ref{fig:fig2}a and Fig.~\ref{fig:fig2}c it is obvious that the intensities of $\Delta F=-2$ group MI transitions at 500 G for $\sigma^-$ polarization is about $10^{5}$ times bigger than for $\sigma^+$ polarization, whereas comparing Fig.~\ref{fig:fig2}b and Fig.~\ref{fig:fig2}d the intensities of the $\Delta F=+2$ group MI transitions at 650 G for $\sigma^+$ polarization are greater about 4 times than for $\sigma^-$ polarization. These two observations lead to a conclusion that there is a definite range of magnetic field when evident circular dichroism exists.

We emphasize the distinctions between $\sigma^+$ and $\sigma^-$ excited MI transitions probabilities in each group by showing the dynamics of the ratios of transitions probabilities depending on the $B$-field in Fig. \ref{fig:fig3}. The red lines are the ratio of the probabilities of the $3^+/1^-$ and $4^-/2^+$ MI transitions for $^{85}$Rb and the blue lines are the ratio of the $7^+/5^-$ and $8^-/6^+$ transitions probabilities for $^{87}$Rb. For convenience, we denote by $A^+$ and $A^-$ the transitions intensities for $\sigma^+$ and $\sigma^-$ excitation, respectively.
While the ratio $A^+/A^-$ of the transitions probabilities for the first group reaches almost five (already demonstrating the existence of dichroism), for the second group the ratio $A^-/A^+$ reaches several hundreds of thousand. Such difference in each group  results in an explicit circular dichroism. The revealed phenomenon can be termed as "frequency-controllable MCD", since by varying the laser field frequency and polarization, it is possible to switch dominating transition intensity from $\sigma^+$ (for $\Delta F = +2$ MI transitions) to $\sigma^-$ (for $\Delta F = -2$ MI transitions).

To describe the dichroism quantitatively, we introduce the figure of merit coefficient $C_{MCD}$, defined as $C_{MCD}= (A^+-A^-) /{ (A^+ + A^-)}$. Magnetic field dependence of $C_{MCD}$ for each group and for $^{85}$Rb (red) and $^{87}$Rb (blue) are plotted in Fig. \ref{fig:fig4}. The solid lines correspond to $\Delta F=+2$ groups and the dashed lines correspond to the $\Delta F=-2$ group. The value $C_{MCD} = 0$ demonstrates the absence of MCD, while $C_{MCD} > 0$ or $C_{MCD} < 0$ indicates the presence of stronger transition intensity for $\sigma^+$ or $\sigma^-$ excitation, respectively. The outermost values $C_{MCD} = \pm1$ correspond to complete suppression of the transition intensity to zero for $\sigma^-$ or $\sigma^+$ excitation, accordingly. The highest negative value $C_{MCD} = -$1 is reached at $B > 80$~G for $F_g=3 \rightarrow F_e=1$ transition of $^{85}$Rb, and at $B > 200~$G for $F_g=2 \rightarrow F_e=0$ transition of $^{87}$Rb, manifesting complete vanishing of $\sigma^+$ MI transition. The highest positive value of $C_{MCD}$ for  $F_g=2 \rightarrow F_e=4$ transition of $^{85}$Rb and $F_g=1 \rightarrow F_e=3$ transition of $^{87}$Rb always stay below $+1$, approaching only to 0.6. We have observed the MI transitions behavior also for Cs and K atoms both experimentally and theoretically. However, we provide the spectra only for natural Rb atoms, to have comparison of different isotopes. According to the mentioned above observation, the highest value of the figure of merit reaches to 0.6 for the $\Delta F = +2$ and $\sim -1$ for the $\Delta F = -2$ groups of transitions of the Cs and K. Thus, we generalize the described phenomena, that is, the strongest MCD occurs for $\Delta F = -2$ transitions of $D_2$ lines and the MI behavior is the same for all bosonic alkali atoms.

As we mentioned, MCD is observed in a particular range of average magnetic fields ($B< B_0$), while in the case of Hyperfine Paschen-Back (HPB) regime ($B\gg B_0$) the MCD is absent. This is due to the decoupling of $I$ and $J$ quantum numbers in strong magnetic fields ($B\gg B_0$), which results in the change of the basis of quantum numbers. In this case, all the MI transitions become forbidden again.

\begin{figure}[t]
\begin{center}
\includegraphics[width=0.5\textwidth]{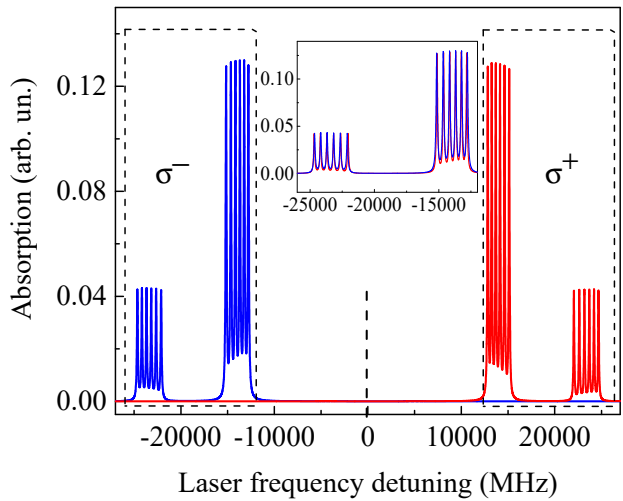}
\caption{\label{fig:fig5} Calculated absorption spectra of $^{85}$Rb $D_2$ line for $B = 1~$T in the case of $\sigma^+$ (red) and $\sigma^-$ (blue) polarized excitation. Zero detuning frequency position corresponds to the weighted center of 5S$_{1/2} \rightarrow 5$P$_{3/2}$ line. The inset illustrates the "mirror" overlap of $\sigma^-$ and $\sigma^+$ spectra when $\sigma^+$ spectrum is flipped horizontally. FWHM is 100~MHz.}
\end{center}
\end{figure}

In contrary to mentioned above asymmetry, to demonstrate the complete symmetry  $D_2$ line absorption spectra of $^{85}$Rb for $B = 1$~T (HPB regime) are calculated in the case of circularly polarized excitation shown in Fig.~\ref{fig:fig5}.  As is clearly seen, the groups of $\sigma^+$ and $\sigma^-$ polarized transitions exhibit mirror symmetry with respect to the zero detuning frequency. This behavior is emphasized in the inset, where the frequency of $\sigma^+$ spectrum is inverted (multiplied by $-1$), demonstrating almost complete overlap of $\sigma^+$ and $\sigma^-$ spectra, the mismatch is only $\sim 0.1~\%$.

In conclusion, we derived a general rule for the intensities of atomic MI transitions for the groups of $\Delta F=+2$ (first) and 
$\Delta F=-2$ (second) of $D_2$ lines of all bosonic alkali atoms (in total $>$ 50 transitions) excited by $\sigma^+$ and $\sigma^-$ circularly polarized radiation in an intermediate external magnetic field. We show that in a particular range of magnetic field transitions intensities in the first group are stronger for the case of $\sigma^+$ laser polarization and vice-versa for the second group. We demonstrate symmetry breaking by dint of complete MCD for the group of $\Delta F=-2$ transitions. In contrast to this asymmetry we also show the full mirror symmetry in HPB regime, when  MCD is absent in strong magnetic fields. Frequency-controllable MCD can be used to switch the dominating transition intensity from $\sigma^+$ to $\sigma^-$ transition by scanning and changing the polarization of the laser field. The simplicity of the experimental setup is demonstrated. The observed strong dichroism (see $\textit{e.g.}$ the sharp peak $A^-/A^+ > 10^{11}$ for $B \approx 255$~G in Fig. \ref{fig:fig3}b) can be of interest as a diagnostic (detection) tool in the experiments on nuclear spin-independent parity violating weak interaction of the electron with the nucleus associated with the weak charge $Q_W$, and in the experiments on nuclear spin-dependent parity non-conservation associated with interaction of the electron with the nuclear anapole moment \cite{gom1,she1}. The elaboration of new experimental schemes and the corresponding estimates are beyond the scope of this paper.

The authors are grateful to D. Budker, A. Gogyan and H. Karapetyan. This work was supported by the State Committee for Science, Ministry of Education and Science of the Republic of Armenia (projects No. 15T-1C040 and 15T-1C277). The research was conducted in the scope of the International Associated Laboratory IRMAS (CNRS-France and SCS-Armenia).

\end{document}